\documentclass[aps,twocolumn,noshowpacs,superscriptaddress,floatfix]{revtex4}

\usepackage[utf8]{inputenc}
\usepackage{braket}
\usepackage{amsmath,amssymb}
\usepackage[justification=justified]{caption}
\usepackage{qcircuit}
\usepackage{xcolor}
\usepackage{verbatim}
\usepackage[normalem]{ulem}
\usepackage[rightcaption]{sidecap}
\usepackage{subfig}

\usepackage[section]{algorithm} 
\usepackage{amsthm,algorithmic,epsfig,tikz}

\def\bra#1{\mathinner{\langle{#1}|}}
\def\ket#1{\mathinner{|{#1}\rangle}}

\renewcommand{\part}[2]{\frac{\partial #1}{\partial #2}}

\usepackage{float}
\usepackage{multirow}

\begin{document}

\title{Benchmarking Amplitude Estimation on a Superconducting Quantum Computer}

\author{Salvatore Certo}
\email{scerto@deloitte.com}
\affiliation{Deloitte Consulting, LLP}

\author{Anh Dung Pham}
\affiliation{Deloitte Consulting, LLP}

\author{Daniel Beaulieu}
\affiliation{Deloitte Consulting, LLP}

\date{\today}

\begin{abstract} 
Amplitude Estimation (AE) is a critical subroutine in many quantum algorithms, allowing for a quadratic speedup in various applications like those involving estimating statistics of various functions as in financial Monte Carlo simulations.  Much work has gone into devising methods to efficiently estimate the amplitude of a quantum state without expensive operations like the Quantum Fourier Transform (QFT), which is especially prohibitive given the constraints of current NISQ devices.  Newer methods have reduced the number of operations required on a quantum computer and are the most promising near-term implementations of the AE subroutine. While it remains to be seen the exact circuit requirements for a quantum advantage in applications relying on AE, it is necessary to continue to benchmark the algorithm's performance on current quantum computers and the circuit costs associated with such subroutines.

Given these considerations, we expand on results from previous experiments in using Maximum Likelihood Estimation (MLE) to approximate the amplitude of a quantum state and provide empirical upper bounds on the current feasible circuit depths for AE on a superconducting quantum computer.  Our results show that MLE using optimally-compiled circuits can currently outperform naive sampling for up to 3 Grover Iterations with a circuit depth of 131, which is higher than reported in other experimental results.  This functional benchmark is one of many that will be continually monitored against current quantum hardware to measure the necessary progress towards quantum advantage.

\end{abstract} 

\maketitle

\section{Introduction}
Many algorithms have been created which can provide a quadratic speedup of certain calculations \cite{montanaro}, including those in finance like Monte Carlo simulations for derivative pricing and Value at Risk (VaR).  These algorithms involve applying functions to one or more random variables and extracting the expected value of the function which is the quantity of interest.  For example, finding the fair price for a European Call Option includes evaluating the payoff function based on the strike price and the distribution of the expected spot prices for the asset at maturity.  In a quantum setting, this example would involve several steps, including loading the spot price distribution into a quantum state, computing the payoff function, and finally extracting the expected value from the amplitude of a quantum state \cite{egger2019credit}\cite{W19}\cite{S19}\cite{O19}.  

While these steps leverage unique properties of quantum mechanics, it is only in the last step of extracting the expected value where the quadratic speedup is obtained.  The retrieval of the expected value of these computations rely on Quantum Amplitude Estimation, which converges as $O(1/M)$ compared to the classical convergence rate of $O(1/\sqrt{M})$\cite{montanaro}.  
Therefore there is a motivation to perform Amplitude Estimation (AE) in an efficient and accurate way, and to benchmark its success on current hardware.  Formally, the goal of AE is to provide an approximation of $a$ prepared by an operator $\mathcal{A}$ such as in 
\begin{equation}
\mathcal{A}\ket{0_n}\ket{0} = \sqrt{1-a}\ket{\psi_0}_n\ket{0} + \sqrt{a}\ket{\psi_1}_n\ket{1}
\end{equation}

In other words, there is some set of quantum operations $\mathcal{A}$ (e.g. the payoff function of a call option) that maps a quantity of interest (e.g. the expected value of the option) into a quantum state $\ket{\psi_1}\ket{1}$ and the goal is to find the amplitude of that state, $\sqrt{a}$.
The canonical method \cite{tapp} was to use Quantum Phase Estimation (QPE) to provide an estimate for the amplitude.  This approach relied on using a controlled version of the grover iterate $\mathcal{Q}$, where $\mathcal{Q}$ is defined as 
\begin{equation*}
\mathcal{Q}= \mathcal{A}\mathcal{S}_0\mathcal{A^\dagger}\mathcal{S}_\psi
\end{equation*}

Here $\mathcal{S}_\psi = \mathcal{I} - 2\ket{\psi_1}_n\bra{\psi_1}_n \bigotimes \ket{1}\bra{1}$ and $\mathcal{S}_0 = \mathcal{I} - 2\ket{0}_{n+1} \bra{0}_{n+1} $.  In other words, $\mathcal{S}_\psi$ adds a negative sign to the $\ket{\psi_1}$ states and $\mathcal{S}_0$ reflects around the zero state.

By then performing Quantum Fourier Transform (QFT) on the control register of size $m$ we would get an estimate for $a$.

\begin{figure}[h]
    \centering
    \includegraphics[width=240px]{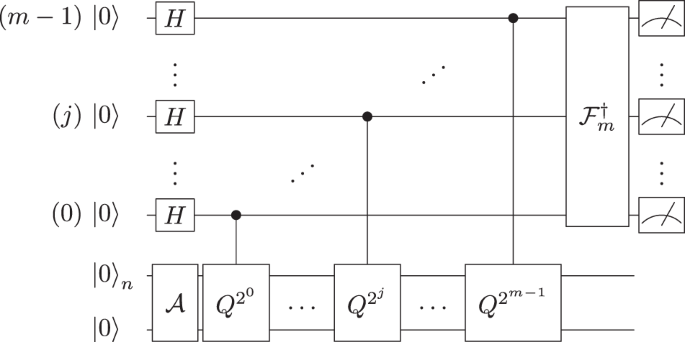}
    \caption{Canonical algorithm for Amplitude Estimation by Quantum Phase Estimation.}
    \label{fig:qpe_ae}
\end{figure}

This method has features that make it unsuitable for current Noisy Intermediate Scale Quantum devices. The qubit topography of many quantum systems make controlled gates expensive and can quickly increase the depth of the circuit.  Similarly, QFT requires expensive operations that are generally not practically implementable in current devices.

There has been active research into other ways to conduct AE, some even involving variational methods \cite{plekhanov2021variational}.  Most current methods involve non-controlled applications of $\mathcal{Q}$ and some classical post processing to create an estimate of $a$ \cite{grinko2019iterative}\cite{S20}.  These methods rely on the fact that $\mathcal{A}\ket{0_n} = \cos{\theta}\ket{\psi_0} + \sin{\theta}{\ket{\psi_1}}$.  If we can get an accurate estimate for $\theta$, then $\sin^2{\theta} = a$.  Included in these methods are the Maximum Likelihood Amplitude Estimation (MLAE) algorithm \cite{S20}, which applies successive applications of the Grover Iterate $\mathcal{Q}$ and finds the $\theta$ that is most likely given the results.  A benefit of MLAE is that it allows us to control the number of applications as well as the number of circuit evaluations for the $\mathcal{Q}$ oracle.

Given a number $m_k$ of repetitions of $\mathcal{Q}$, the number of circuit evaluations $N_k$, and the number of times $h_k$ that $\ket{\psi_1}\ket{1}$ was measured, the probability of measuring $\ket{\psi_1}\ket{1}$ is $P(m_k;a) = \sin^2{((2m_k +1)}\theta_a)$.  The likelihood function is therefore

\begin{equation}
\mathcal{L}(h;a)= \prod_{k=0}^M [P(m_k;a)]^{h_k}[1-P(m_k;a)]^{N_k-h_k}
\end{equation}

The number of repetitions of $\mathcal{Q}$ as well as the number of circuit evaluations per each repetition define the sequencing schedule for the algorithm.  Different schedules have been proposed, including the Linearly Incremental Sequence (LIS), the Exponential Incremental Sequence (EIS), and the Power Law schedule.  The experiments we conducted used the simple LIS schedule.

\section{Motivation for Experiment}

Experimental results were recently published \cite{lowdepth} that showed evidence of amplitude estimation outperforming naive sampling and low error rates on a state-of-the-art trapped ion quantum computer.  It is known that the different types of quantum computers have advantages and disadvantages, and while some work has been published on benchmarking AE on superconducting hardware \cite{rao2020quantum}, there has been little published on actual circuit depths and their effects on amplitude estimation.  While quantum engineers use high level languages to construct circuits, they are first compiled to solve the many constraints a particular quantum computer has, including qubit topography and actual basis gate implementations.  This can have dramatic effects on the actual circuit depths, number and types of gates used, thereby altering the performance of the algorithm than what may be expected based on the non-compiled circuit.  Therefore it is critical to continue to evaluate the status of current computers as well as the effect of circuit depth on their performance, on which we report in this paper.

\section{Amplitude Estimation Circuits used in MLE Algorithm}

One can construct any arbitrary circuit to use for amplitude estimation, the only requirement being that there exists a well-defined quantum state from which we can benchmark the results.  For our purposes, we wanted to design an experiment that is reproduce-able, scalable, and has advantageous properties related to the algorithm we are benchmarking.  The circuits used in \cite{lowdepth} share some of these properties.  They originally were proposed to be used as a unary data loader, a method to load unit vectors and matrices into a quantum system.  As it pertains to amplitude estimation and shown in \cite{lowdepth}, it also has nice properties for both the oracle $\mathcal{S}_\psi$ and diffuser $\mathcal{A}\mathcal{S}_0\mathcal{A^\dagger}$ of the Grover Iterate, requiring single Z gates for both $\mathcal{S}_\psi$ and $\mathcal{S}_0$.

The experiments previously mentioned used unary data loaders to load two unit vectors and compute their inner product in the amplitude of a quantum state that is to be estimated.  It relied on the $\mathcal{RBS}$ gate which is a two qubit unitary matrix that is described as:
\begin{equation}
    \mathcal{RBS}(\theta) = 
    \begin{pmatrix}
1 & 0 & 0 & 0\\
0 & \cos{\theta} & \sin{\theta} & 0\\
0 & -\sin{\theta} & \cos{\theta} & 0\\
0 & 0 & 0 & 1
\end{pmatrix}
\end{equation}

Based on the unit vectors, there is a method for finding $\theta$.  While we refer to \cite{lowdepth} and \cite{johri2020nearest} for more information on constructing the specific gates, we will briefly provide an example of using them.  Given two unit vectors, $\ket{a}$ and $\ket{b}$, we can find angles through the procedure in \cite{johri2020nearest} to construct the circuit below that will compute the square of their inner product $\braket{a|b}$ in the amplitude of the first qubit.  

\begin{figure}[h]
    \centering
    \includegraphics[width=240px]{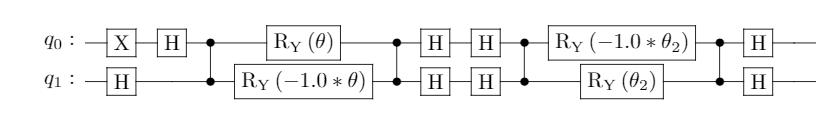}
    \caption{Circuit for computing the inner product between two 2 dimensional unit vectors.}
    \label{fig:rbs}
\end{figure}

For instance, \newline

$\ket{a} = \begin{bmatrix} -0.96184207 & 0.27360523 \end{bmatrix}$

$\ket{b} = \begin{bmatrix} 0.96845237 & 0.24919873 \end{bmatrix}$

$\braket{a|b}^2 = \ket{01} = .745314789774813$ \newline

We used similar initial gate configurations in Qiskit as \cite{lowdepth}, however both the hardware we ran our circuits on as well as the compilers changed the gate configurations.  Still, shown in Figure 3 and Figure 4 are the initial circuits for the 4 dimensional inner product circuits which represent $\mathcal{A}$ and $\mathcal{A}\mathcal{S}_0\mathcal{A^\dagger}\mathcal{S}_\psi\mathcal{A}$.

\begin{figure}[h]
    \centering
    \includegraphics[width=240px]{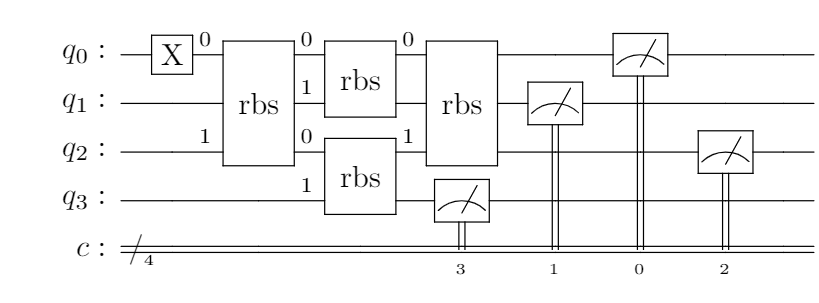}
    \caption{Inner product estimation circuit for $\mathcal{A}$}
    \label{fig:rbs}
\end{figure}

\begin{figure}[h]
    \centering
    \includegraphics[width=230px]{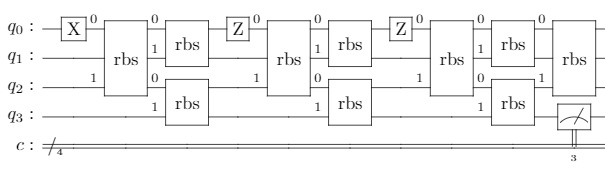}
    \caption{Circuit for a single grover iteration of $\mathcal{A}\mathcal{S}_0\mathcal{A^\dagger}\mathcal{S}_\psi\mathcal{A}$}
    \label{fig:rbs}
\end{figure}

In the 4 dimensional case, $\mathcal{A}$ will load the states $a\ket{0001} + b\ket{0010} + c\ket{0100} + d\ket{1000}$ where $a,b,c,d$ are the amplitudes and $a^2+b^2+c^2+d^2 = 1$.  We are interested in the amplitude of $a$, the probability of the $\ket{0001}$ state which represents the squared inner product of the 4 dimensional unit vectors.

While we tested a trivial example on 2 qubits, the main result of our work was in estimating the inner product of 2 random 4 dimensional unit vectors.  In this case, we constructed $\mathcal{A}$ and $\mathcal{A}\mathcal{S}_0\mathcal{A^\dagger}\mathcal{S}_\psi\mathcal{A}$ as shown above.  The quantum computer we ran our circuits on took the high level circuits and compiled them to the allowable basis gates.  We then used Qiskit and Pytket to optimize these compiled circuits as aggressively as allowed.  

To estimate the amplitude of the first qubit which represents the inner product of the unit vectors, we added Grover Iterates from $(0...7)$ where $0$ is just $\mathcal{A}$.  Once compiled based on the 3 different methodologies, we ran the circuit 500 times and used the results as input to our Maximum Likelihood Estimation algorithm.  Therefore by using $m_k = 7$ of repetitions of $\mathcal{Q}$, we also used the previous results of $m_k = (0...6)$ for our calculation.  

We generated random unit vectors whose inner product is uniformly distributed and instantiated our initial distribution for $\theta$ in the MLE algorithm to be a uniform distribution between $[0,\frac{\pi}{2}]$.  For each set of unit vector pairs, we compared the MLE output and traditional sampling method to the exact inner product, giving us our estimate of relative error.

\section{IBM Nairobi Specifications}

The experiments conducted were done using the IBM Nairobi superconducting computer.  It has 7 qubits with
a quantum volume of 32.  The exact specifications are detailed in the specifications table.


\begin{figure}[h]
    \centering
    \includegraphics[height=160px]{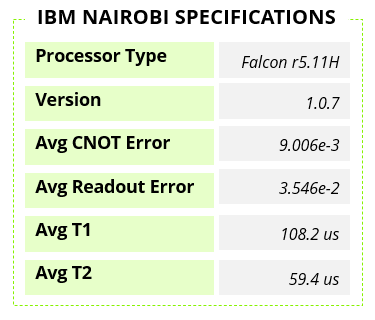}
    \label{fig:nairobi}
\end{figure}

\begin{figure*}[h]
    \centering
    \includegraphics[height = 300px]{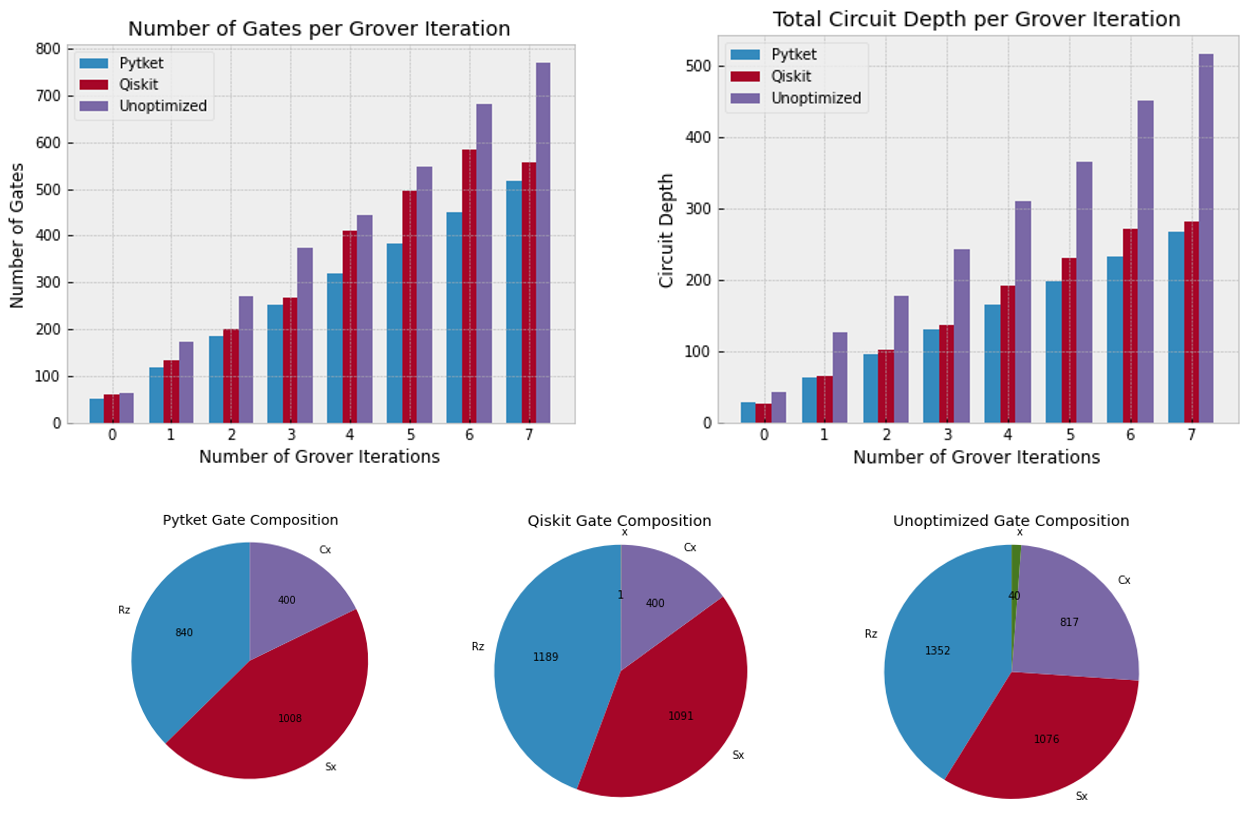}
    \caption{Depth, gate count, and gate composition of various compilation methods}
    \label{fig:nairobi}
\end{figure*}

\section{Comparison of Optimized Circuits}

\begin{figure*}[h]
    \centering
    \includegraphics[width=500px]{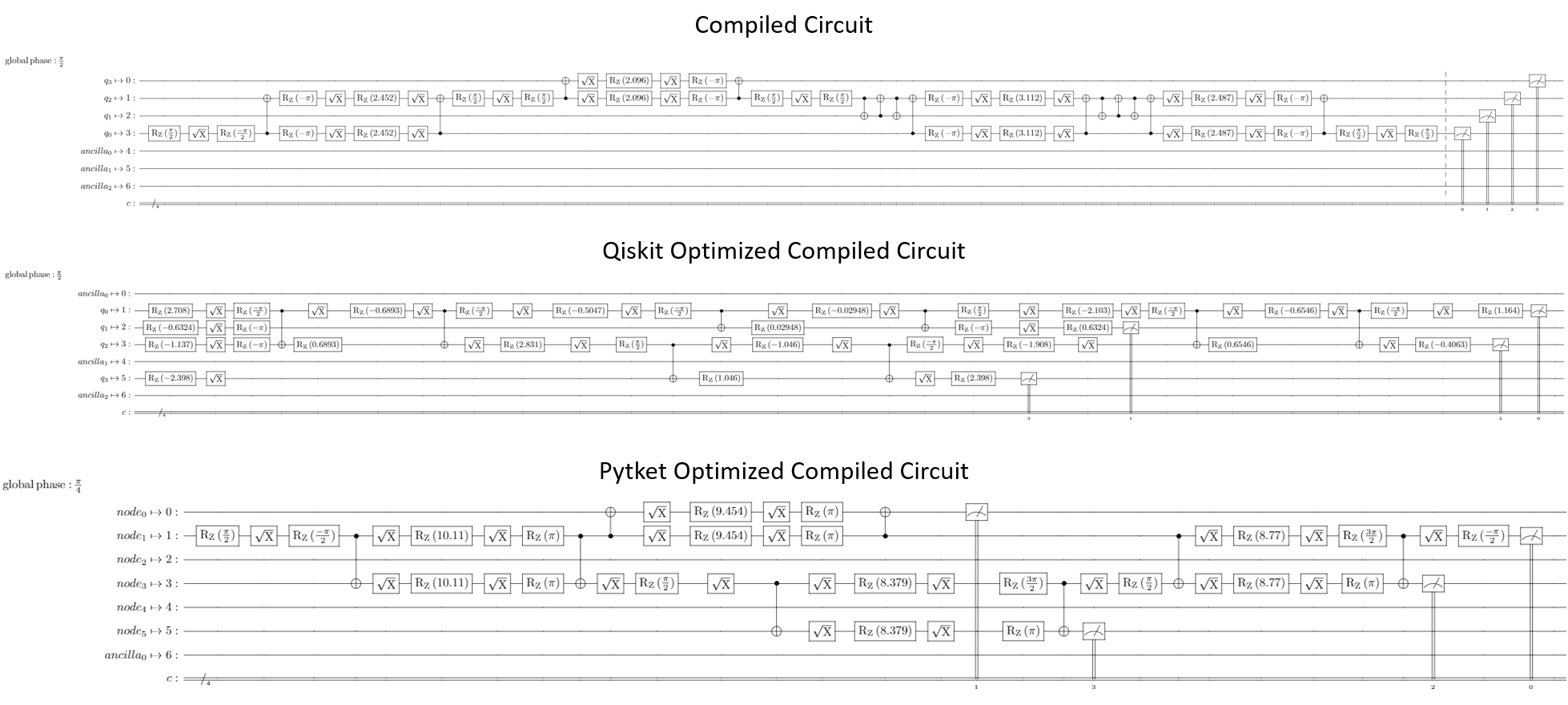}
    \caption{Comparison of $\mathcal{A}$ circuits.}
    \label{fig:nairobi}
\end{figure*}

The RBS gates outlined above have significant depths when compiled on IBM hardware.  In fact, they can reach a depth of over 500 for 7 Grover Iterations.  Knowing the need to reduce these depths and optimize the circuits as much as possible, we explored two well known open source compilers provided by Qiskit \cite{Qiskit} and Pytket \cite{pytket}.  There are many ways to optimally compile a quantum circuit, none of which are trivial.  Considerations need to be made for the goal of the algorithm, the specific hardware, and the types of gates required along with their typical error rates.  

Both Qiskit and Pytket have a similar set of compilation passes, first translating the high level circuit into the required basis gates of the hardware, then attempting to optimize the circuit to reduce depth and noise.  As referenced in \cite{pytket}, Pytket's optimization compiler falls into two broad categories: peephole optimizations and macroscopic optimizations.  Peephole optimizations are similar to those in classical settings, where a sliding window looks for local patterns that can be substituted with equivalent sub-circuits with lower gate counts or depth.  Macroscopic optimization looks for non-local structures that can be synthesized to increase parallelization  or fewer gates, potentially leading to another pass of peephole optimization.  We refer to \cite{pytket} for more detailed information.

In figure 6 we show a sample of circuits for certain 4 dimensional vectors with no Grover iterations.  Both Qiskit and Pytket substantially reduced the two-qubit interactions, gate counts, and overall circuit depth.  In figure 5 we detail the number of gates and circuit depth per Grover iteration, as well as the total (sum of all 7 sub-circuits) counts per gate type.  We can see that both Qiskit and Pytket reduced the number of gates and the circuit depth; the circuit depth was reduced by almost half and was a substantial improvement over the standard compilation procedure.  

Furthermore, both found the minimum number of CNOT gates that could be used, while the standard compiler used more than double. We leave it to further experiments to benchmark the different ways to optimize quantum circuits; the results we present are not direct comparisons between either approaches.  If we were to do so, we would choose a variety of metrics, algorithms, and circuit types to quantitatively measure them.  Here we are simply showing in this specific case both are worthwhile to use versus the standard compiler.

\section{Minimal Depth Amplitude Estimation Results}

Here we will review the results of estimating the amplitude of random unit vectors for both 2 and 4 dimensions.  We used a linear sequencing schedule for MLE, of 500 shots each up to 7 Grover Iterations.  The experiment was repeated 50 times each, and the results averaged.  For naive sampling, we ran the circuits for up to 3000 shots in increments of 500.  

For the 2 qubit experiments, both the ordinarily-compiled circuits as well as the optimized circuits outperformed naive sampling.  Interestingly, the larger number of shots for naive sampling did not reduce the error, which is most likely due to the fact that the inherent noise found on current hardware prohibits better amplitude estimates.  As shown in Figure 7, The lowest recorded mean error of .0017 was after 7 Grover iterations, although after 3 Grover iterations the decrease in error became minimal.  

While the 2 qubit experiment is trivially small, it demonstrated that MLE does outperform simple averaging of samples and after even 1 Grover iteration the error decreased by half.  It also showed the impact of reducing circuit depth, complexity, and optimizing it for a particular quantum computer.  

\begin{figure}[!h]
    \centering
    \includegraphics[height=180px]{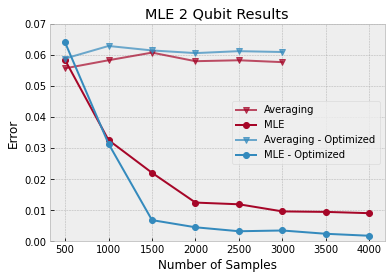}
    \caption{2 qubit AE results for ordinary transpiled and optimized depth circuits.  The circles represent the maximum number of Grover Iterations used in MLE, with 500 shots per each iteration.}
    \label{fig:2qubits}
\end{figure}

\begin{figure}[h!]
    \centering
    \includegraphics[width=230px]{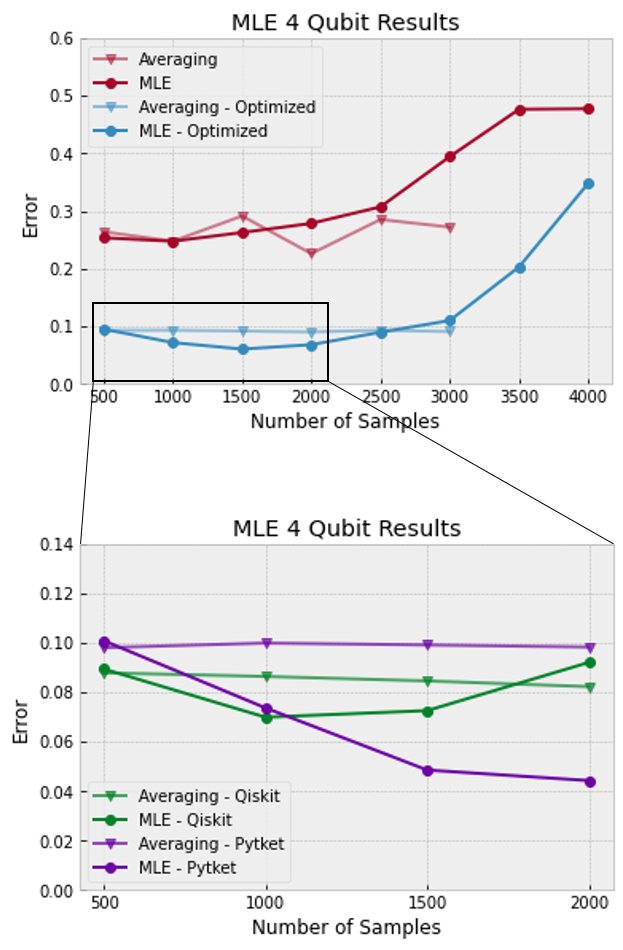}
    \caption{4 qubit AE results for ordinary compiled and optimized depth circuits.  The optimized circuits were done using level 2 optimization provided by Pytket and level 3 optimization by Qiskit. The best results achieved in our experiments were found after 3 Grover iterations on the Pytket optimally-compiled circuits, achieving a mean error of .0442.}
    \label{fig:nairobi_4}
\end{figure}

The 4 qubit results are more significant as the experiment still showed MLE outperforming naive sampling with much larger circuit depths and 2-qubit gates.  The circuit compiler provided by both Qiskit and Pytket reduced the error rates for both simple averaging and MLAE, with Pytket providing a shallower circuit with lower mean error.  Even with an optimal circuit layout, the depths reached a maximum of around 300 and the errors did deteriorate after 3 Grover iterations caused by decoherence of the qubits and compounded errors throughout the circuit.  Both Qiskit and Pytket have different levels of circuit compilers, ranging from (0,1,2,3) for Qiskit and (0,1,2) for Pytket.  Each of these compilation levels differ in their aggressiveness for optimizing the ciruits and reducing the depth.  We used the most aggressive compilers from each and the results are shown in Figure 8.  

It is worthy to note that while Pytket seemed more robust to noise in the higher number of Grover Iterations, this could be the result of several causes.  We used 50 random vectors for each experiment, and so the exact distribution of this small sample size could account for the discrepancy.  Qiskit had a lower sampling error, which we also believe could have been the result of noise in the quantum system as well as the random vectors.  Nonetheless, both offer substantial improvement over non-optimized circuits and show how significant an impact a well-compiled circuit performs over an ordinarily-compiled one.  Even the results from the naive averaging of the non-optimized circuits had a much larger variance than the optimized circuits from both Qiskit and Pytket, showing that the compilation provided by both substantially reduces unwanted noise and stabilizes the results.

Compared to \cite{lowdepth}, our optimized circuits had higher depths than the experiments performed on a trapped ion computer, which reported two-qubit gates up to ninety two and a maximum circuit depth of sixty two.  The depths reported here are much larger compared to previous experiments as we were able to achieve a minimum mean error of around .04 with circuit depths of 131 for 3 Grover iterations. In fact, while the circuits constructed were minimal with respect to IBM's superconducting computers, they are some of the deepest circuits reported thus far that still obtained a significant reduction in error while performing MLAE.  That said, recent work has determined that the threshold for quantum advantage in certain Monte Carlo simulations requires a t-depth of around 54 million \cite{threshold}, and so while current hardware is out of reach for such performance, we hope that functional benchmark tests such as this experiment continue to evaluate the maturity of current systems in the pursuit of such an advantage.



\section{Conclusion}

In this paper we have expanded on previous work \cite{lowdepth} in using circuits that compute the inner product between unit vectors to benchmark the Maximum Likelihood Amplitude Estimation algorithm.  While we have shown that both the circuit optimizers by Qiskit and Pytket work very well in reducing noise and increasing the allowable number of Grover iterations we can perform in the algorithm, we have more importantly provided an empirical upper bound on the maximum allowable depth for AE on current hardware, a metric that will be will closely monitored as the industry progresses. \newline

\section{Disclaimer}

About Deloitte: Deloitte refers to one or more of Deloitte Touche Tohmatsu Limited (“DTTL”), its global network of member firms, and their related entities (collectively, the “Deloitte organization”). DTTL (also referred to as “Deloitte Global”) and each of its member firms and related entities are legally separate and independent entities, which cannot obligate or bind each other in respect of third parties. DTTL and each DTTL member firm and related entity is liable only for its own acts and omissions, and not those of each other. DTTL does not provide services to clients. Please see www.deloitte.com/about to learn more.

Deloitte is a leading global provider of audit and assurance, consulting, financial advisory, risk advisory, tax and related services. Our global network of member firms and related entities in more than 150 countries and territories (collectively, the “Deloitte organization”) serves four out of five Fortune Global 500® companies. Learn how Deloitte’s
approximately 330,000 people make an impact that matters at www.deloitte.com. 
This communication contains general information only, and none of Deloitte Touche Tohmatsu Limited (“DTTL”), its global network of member firms or their related entities (collectively, the “Deloitte organization”) is, by means of this communication, rendering professional advice or services. Before making any decision or taking any action that
may affect your finances or your business, you should consult a qualified professional adviser. No representations, warranties or undertakings (express or implied) are given as to the accuracy or completeness of the information in this communication, and none of DTTL, its member firms, related entities, employees or agents shall be liable or
responsible for any loss or damage whatsoever arising directly or indirectly in connection with any person relying on this communication. 
Copyright © 2021. For information contact Deloitte Global.

\bibliographystyle{plain} 
\bibliography{b1.bib}

\begin{thebibliography}{10}

\bibitem{Qiskit}
Qiskit: An open-source framework for quantum computing, 2021.

\bibitem{tapp}
Gilles Brassard, Peter Høyer, Michele Mosca, and Alain Tapp.
\newblock Quantum amplitude amplification and estimation.
\newblock {\em Quantum Computation and Information}, page 53–74, 2002.

\bibitem{threshold}
Shouvanik Chakrabarti, Rajiv Krishnakumar, Guglielmo Mazzola, Nikitas
  Stamatopoulos, Stefan Woerner, and William~J. Zeng.
\newblock A threshold for quantum advantage in derivative pricing.
\newblock {\em Quantum}, 5:463, Jun 2021.

\bibitem{egger2019credit}
Daniel~J Egger, Ricardo~Garc{\'\i}a Guti{\'e}rrez, Jordie~Cahue Mestre, and
  Stefan Woerner.
\newblock Credit risk analysis using quantum computers.
\newblock {\em IEEE Transactions on Computers}, 2020.

\bibitem{lowdepth}
Tudor Giurgica-Tiron, Sonika Johri, Iordanis Kerenidis, Jason Nguyen, Neal
  Pisenti, Anupam Prakash, Ksenia Sosnova, Ken Wright, and William Zeng.
\newblock Low depth amplitude estimation on a trapped ion quantum computer,
  2021.

\bibitem{grinko2019iterative}
Dmitry Grinko, Julien Gacon, Christa Zoufal, and Stefan Woerner.
\newblock Iterative quantum amplitude estimation.
\newblock {\em npj Quantum Information}, 7(1):1--6, 2021.

\bibitem{johri2020nearest}
Sonika Johri, Shantanu Debnath, Avinash Mocherla, Alexandros Singh, Anupam
  Prakash, Jungsang Kim, and Iordanis Kerenidis.
\newblock Nearest centroid classification on a trapped ion quantum computer,
  2020.

\bibitem{montanaro}
Ashley Montanaro.
\newblock Quantum speedup of monte carlo methods.
\newblock {\em Proceedings of the Royal Society A: Mathematical, Physical and
  Engineering Sciences}, 471(2181):20150301, Sep 2015.

\bibitem{O19}
Rom{\'a}n Or{\'u}s, Samuel Mugel, and Enrique Lizaso.
\newblock Quantum computing for finance: overview and prospects.
\newblock {\em Reviews in Physics}, page 100028, 2019.

\bibitem{plekhanov2021variational}
Kirill Plekhanov, Matthias Rosenkranz, Mattia Fiorentini, and Michael Lubasch.
\newblock Variational quantum amplitude estimation, 2021.

\bibitem{rao2020quantum}
Pooja Rao, Kwangmin Yu, Hyunkyung Lim, Dasol Jin, and Deokkyu Choi.
\newblock Quantum amplitude estimation algorithms on ibm quantum devices, 2020.

\bibitem{pytket}
Seyon Sivarajah, Silas Dilkes, Alexander Cowtan, Will Simmons, Alec Edgington,
  and Ross Duncan.
\newblock t|ket⟩: a retargetable compiler for nisq devices.
\newblock {\em Quantum Science and Technology}, 6(1):014003, Nov 2020.

\bibitem{S19}
Nikitas Stamatopoulos, Daniel~J Egger, Yue Sun, Christa Zoufal, Raban Iten,
  Ning Shen, and Stefan Woerner.
\newblock Option pricing using quantum computers.
\newblock {\em Quantum}, 4:291, 2020.

\bibitem{S20}
Yohichi Suzuki, Shumpei Uno, Rudy Raymond, Tomoki Tanaka, Tamiya Onodera, and
  Naoki Yamamoto.
\newblock Amplitude estimation without phase estimation.
\newblock {\em Quantum Information Processing}, 19(2):75, 2020.

\bibitem{W19}
Stefan Woerner and Daniel~J Egger.
\newblock Quantum risk analysis.
\newblock {\em npj Quantum Information}, 5(1):1--8, 2019.

\end{thebibliography}

\end{document}